\documentclass[12pt]{article}
\usepackage{amssymb}
\usepackage{amsmath}
\usepackage{epsfig}

\begin{document}

\title{\hspace{4.1in}{\small CERN-TH/2003-146}\bigskip \\
Neutrino properties from Yukawa structure.}
\author{A. Ibarra$^{a}$ and G. G. Ross$^{a,b}$ \\
$^{a}$ Theory Group, CERN, 1211 Geneva 23, Switzerland\\
$^{b}$Department of Physics, Theoretical Physics, University of Oxford,\\
1 Keble Road, Oxford OX1 3NP, U.K.}
\date{}
\maketitle

\begin{abstract}
We discuss the implications for lepton mixing and CP\ violation of structure
in the lepton mass matrices, for the case that neutrino masses are generated
by the see-saw mechanism with an hierarchical structure for the Majorana
masses. For a particularly interesting case with enhanced symmetry in which
the lepton Dirac mass matrices are related to those in the quark sector, the
CHOOZ angle is near the present limit and the CP\ violating phase relevant
to thermal leptogenesis and to $\nu 0 \beta \beta $ decay is near maximal.
\end{abstract}

\section{Introduction}

The origin of the structure observed in quark and lepton masses and mixing
angles remains one of the most pressing and interesting questions left
unanswered by the Standard Model. The continuing improvement in the
measurement of the CKM and MNS matrix elements and the neutrino masses has
stimulated a renewed theoretical effort to answer these questions.

In the case of quarks one proposed structure going beyond the Standard Model
has proved to be robust, giving a quantitatively accurate prediction for the
Cabbibo angle (strictly $V_{12}^{CKM}$). It follows from the postulate that
the up and down quark mass matrices have a simultaneous \textquotedblleft
texture\textquotedblright\ zero in the $(1,1)$ position\footnote{%
A texture zero does not imply a matrix element is absolutely zero, but only
that it is small enough so that it does not significantly affect the masses
and mixing angles.} and that the magnitude of the matrix elements are
symmetric for the first two generations\cite{gsto}. The measured masses and
mixing angles are consistent with additional texture zeros\cite%
{Ramond:1993kv}, although this may require a departure from the symmetric
form of the mass matrices\cite{Roberts:2001zy}. One reason for the interest
in texture zeros is that they may indicate the presence of a new family
symmetry which require certain matrix elements be anomalously small. Thus
identification of texture zeros may be an important step in unravelling the
origin of the fermion masses and mixings.

In this paper we extend the analysis of possible texture zeros to the lepton
sector for the case that neutrino masses are given by the see-saw mechanism%
\cite{seesaw}. In analogy with the quark case we consider the predictions
resulting from a symmetric form for the magnitudes of the Dirac mass matrix
elements together with texture zeros. Of particular interest is the case of
simultaneous zeros in the $(1,1)$ position. If this proves to be the case it
would be a strong indication of a symmetry between the up and the down
quarks and the charged lepton and neutrino sectors respectively. For the
case that the Majorana mass matrix does not contribute significantly to
lepton mixing we obtain predictions for the CHOOZ mixing angle and for the
CP\ violating phases. If the neutrino Majorana mass does contribute
significantly to mixing these predictions may be viewed as indicative to the
magnitude of these parameters barring what would seem to be an unnatural
cancellation between the contribution of the Dirac and Majorana sectors. We
also consider the implications further restrictions on the form of the
lepton mass matrices. The analysis is done in the context that the mass of
one of the Majorana neutrinos is anomalously large \cite{rw}. This case
includes the possibilities that there is sequential right hand neutrino
dominance\cite{King:1998jw} that offers an attractive way of explaining near
bi-maximal neutrino mixing in the case that the quark and neutrino Dirac
mass matrices are related\cite{King:2001uz},\cite{Ross:2002fb}.

The paper is organised as follows. In Section \ref{param} we review a
general parameterisation for the effective light neutrino masses for the
case of the see-saw mechanism that is useful in studying the implications of
texture zeros. We discuss the constraints on this parameterisation coming
from texture zeros, from a symmetric form of the magnitudes of the mass
matrix elements and from the case that one of the Majorana neutrinos is
anomalously large. In Section \ref{results} we apply this parameterisation
to derive general constraints on neutrino mixing and CP\ violation and
consider the implications for leptogenesis. Section \ref{summary} summarizes
the results.

\section{Parameterisation of the see-saw mechanism\label{param}}

We consider the case of three generations of left-handed $SU(2)$ doublet
neutrinos, $\nu _{L,i,}$ and three generations of right-handed Standard
Model singlet neutrinos, $\nu _{R,i}$. The Lagrangian responsible for lepton
masses has the form%
\begin{equation}
L_{Mass}^{l}=\nu _{R}^{cT}.Y_{\nu }^{D}.\nu _{L}\langle H^{0}\rangle
+l_{R}^{cT}Y_{l}^{D}.l_{L}\langle \overline{H}^{0}\rangle -\frac{1}{2}\nu
_{R}^{cT}.M_{\nu }^{M}.\nu _{R}^{c}  \label{mass}
\end{equation}%
where $Y_{\nu }^{D},$ $Y_{l}^{D}$ are the matrices of Yukawa couplings which
give rise to the neutrino and charged lepton Dirac mass matrices
respectively and $M_{\nu }^{M}$ is the neutrino Majorana mass matrix. We are
interested in studying the implications of simultaneous zeros in $Y_{\nu
}^{D}$ and $Y_{l}^{D}$ for observable quantities, masses and mixing angles
and CP\ violating phases. For the case of quarks and charged leptons it is
easy to do this because the Yukawa couplings are directly related to the
mass matrices. For neutrinos, however, the existence of the Majorana masses
complicates the connection between the Dirac Yukawa couplings and the
neutrino observables. The light neutrino mass matrix, $\mathcal{M}\mathit{,}$
is given by the see-saw form%
\begin{equation}
\mathcal{M}\text{=}Y_{\nu }^{DT}.M_{\nu }^{M-1}.Y_{\nu }^{D}  \label{seesaw}
\end{equation}%
Sometimes it is convenient to use an alternative form for the see-saw
formula, expressing $Y_{\nu }^{D}$ in terms of the neutrino mass
eigenvalues, mixing angles and CP\ violation\cite{Casas:2001sr}. In the
basis in which the Majorana mass matrix, $M_{\nu }^{M},$ is diagonal the
parameterisation has the form%
\begin{equation}
Y_{\nu }^{D}=D_{\sqrt{M}}.R.D_{\sqrt{m}}.W^{\dag }/\langle H^{0}\rangle
\label{ci}
\end{equation}%
where $D_{\sqrt{M}}$ is the diagonal matrix of the square roots of the
eigenvalues of $M_{\nu }^{M}$, $D_{\sqrt{m}}$ is the diagonal matrix of the
roots of the physical masses, $m_{i},$ of the light neutrinos, $W$ is the
neutrino mixing matrix, and $R$ is an orthogonal matrix which parameterises
the residual freedom in $Y_{\nu }^{D}$ once the other parameters are fixed.
It is parameterised by 3 complex \textquotedblleft mixing\textquotedblright\
angles\footnote{%
Up to reflections which can be absorbed in the unknown phases discussed
below.}.

\begin{equation}
R=\left( 
\begin{array}{ccc}
\sin \theta _{2}\sin \theta _{3} & \cos \theta _{1}\cos \theta _{3}+\sin
\theta _{1}\cos \theta _{2}\sin \theta _{3} & \sin \theta _{1}\cos \theta
_{3}-\cos \theta _{1}\sin \theta _{3} \\ 
\sin \theta _{2}\cos \theta _{3} & -\cos \theta _{1}\sin \theta _{3}+\sin
\theta _{1}\cos \cos \theta _{3} & -\sin \theta _{1}\sin \theta _{3}-\cos
\cos \theta _{3} \\ 
\cos \theta _{2} & \sin \theta _{1}\sin \theta _{2} & -\cos \theta _{1}\sin
\theta _{2}%
\end{array}%
\right)  \label{r}
\end{equation}%
where $\theta _{1},$ $\theta _{2},$ $\theta _{3}$ are arbitrary complex
angles. These, together with the three Majorana masses, the three light
neutrino masses, the three mixing angles and three phases of $W$ make up the
eighteen real parameters needed to specify $Y_{\nu }^{D}.$ With this form it
is straightforward to study the implications of a zero in $Y_{\nu }^{D}$ for
the physical measureables.

In our study of texture zeros we will be interested in simultaneous texture
zeros in $Y_{\nu }^{D}$ and $Y_{l}^{D}$. Of course this is basis dependent
as a zero in one basis will not in general remain zero after a rotation. In
this sense the appearance of simultaneous texture zeros specifies the
\textquotedblleft texture zero\textquotedblright\ basis. The idea is that
there is some dynamical reason, such as a family symmetry, which generates
the texture zero structure. For the case of a family symmetry the
\textquotedblleft texture zero\textquotedblright\ basis is just the current
quark basis, defined as the one in which the fermion states are eigenstates
of the family symmetry group. In the phenomenological analysis of texture
zeros this basis choice is taken into account by modifying the
parameterisation so that the charged lepton mass matrix is not diagonal. In
this case it is the combination $U_{l}^{\dagger }W$ that should be
identified with the $MNS$ matrix, where $U_{l}$ is the unitary matrix needed
to diagonalise the charged lepton mass matrix, starting from the texture
basis.

It is instructive to determine how many free parameters are left in $R$ when 
$Y_{\nu }^{D}$ is constrained in various ways. If any element of $Y_{\nu
}^{D}$ is zero, there is a reduction of two complex parameters needed to
specify $Y_{\nu }^{D}$ and a corresponding reduction of the parameters in $%
R. $ For more than $3$ texture zeros there will be relations between
measureable quantities\footnote{%
We include the Majorana mass eigenvalues amongst our \textquotedblleft
measureables\textquotedblright\ and also the mixing angles in $W$; of course
it is necessary to discuss the lepton sector to relate $W$ to $U_{MNS}$.}.
However, depending on the position of the texture zero, there may be
predictions for fewer texture zeros.

For the case that $Y_{\nu }^{D}$ is symmetric (or hermitian or has off
diagonal elements antisymmetric) the number of real parameters needed to
specify it are reduced to $12$, so in this case $R$ is completely
determined. This does not lead to any relations between measurable
quantities but if, in addition, there is a texture zero there will be such
relations (this is the analogue to the GST\ relation in the quark sector).

For the case one of the Majorana masses, $M_{\nu ,3}^{M},$ is anomalously
heavy the Standard Model singlet component, $\nu _{R,3},$does not play a
role in determining the two heaviest of the light neutrino eigenstates.
Following from eq(\ref{seesaw}) we see that in this case the couplings $%
\left( Y_{\nu }^{D}\right) _{3j},j=1..3$ do not contribute to the light
masses and mixing angles. There is also a reduction in the number of
parameters needed to specify $R.$Following from the condition that $Y_{\nu
}^{D}W$ is finite as $M_{\nu ,3}^{M}\rightarrow \infty ,$ we see that in
this limit $R_{3j}\propto \sqrt{1/M_{\nu ,3}^{M}},$ $j=2,3$ and $R_{ij}\leq
O(1),$ $i,j=1..3$ $.$ Inserting these constraints in eq(\ref{r}) we find the
form of $R$ is given by%
\begin{equation}
R=\left( 
\begin{array}{ccc}
\propto \sqrt{1/M_{\nu ,3}^{M}} & \cos z & \pm \sin z \\ 
\propto \sqrt{1/M_{\nu ,3}^{M}} & -\sin z & \pm \cos z \\ 
\sim 1 & \propto \sqrt{1/M_{\nu ,3}^{M}} & \propto \sqrt{1/M_{\nu ,3}^{M}}%
\end{array}%
\right)  \label{srhd}
\end{equation}%
where $z=\theta _{3}-\theta _{1}.$ This $\pm $ refer to a reflection
ambiguity. In practice we can work with the positive sign only and absorb
this ambiguity in the unknown phases specified below. The Yukawa couplings $%
\left( Y_{\nu }^{D}\right) _{ij},$ $i=1,2,$ $j=1..3$ are thus given in terms
of $z$ alone in the limit $M_{\nu ,3}^{M}\rightarrow \infty .$ If we require
the $(1,2)$ block be symmetric, antisymmetric or hermitian, $z$ will be
determined and for $1$ texture zero there will be relations between
measureables. Alternatively more than $1$ texture zero will give relations
even if the $(1,2)$ and $(2,1)$ matrix elements are not related.

\section{The charged lepton mass matrix}

The $MNS$ matrix is given by $U_{l}^{\dagger }W$ and has a contribution
coming from the matrix $U_{l}$ which diagonalises the charged lepton mass
matrix. The latter has to\ reproduce the hierarchical structure of lepton
masses and this may place constraints on the magnitude of the charged lepton
mixing angles. Let us consider the case the lepton mass matrix is symmetric
and that, like the quarks, the hierarchy of lepton masses is due to an
hierarchical structure in the matrix elements and not due to a cancellation
between different contributions. This is what is expected if there is an
underlying Grand Unified symmetry relating leptons to quarks. Moreover a
cancellation between different contributions to lepton masses seems very
difficult to reconcile with an underlying family symmetry as it requires
non-trivial relations between different matrix elements which are difficult
to arrange even in the context of non-Abelian family symmetry. With this
constraint it is easy to limit $\left( U_{l}\right) _{23},$ because $\left(
M_{l}\right) _{23}^{2}\leq m_{\mu }m_{\tau }$, giving 
\begin{equation}
\left\vert \left( U_{l}\right) _{23}\right\vert \leq \sqrt{\frac{m_{\mu }}{%
m_{\tau }}}.  \label{l1}
\end{equation}%
Similarly one obtains a bound on $\left( U_{l}\right) _{12}$ from the
constraint that $\left( M_{l}\right) _{12}^{2}<m_{e}m_{\mu }$ which follows
from taking the deteminant of the mass matrix. This in turn implies 
\begin{equation}
\left\vert \left( U_{l}\right) _{12}\right\vert \leq \sqrt{\frac{m_{e}}{%
m_{\mu }}}  \label{l2}
\end{equation}%
with equality occurring if there is a texture zero in the $(1,1)$ position.

The constraint on $\left( M_{l}\right) _{12}^{2}$ also leads to the
constraint $\left\vert \left( U_{l}\right) _{13}\left( U_{l}\right)
_{23}\right\vert \leq \frac{\sqrt{m_{e}m_{\mu }}}{m\tau }$. If $\left\vert
\left( U_{l}\right) _{23}\right\vert =\sqrt{\frac{m_{\mu }}{m_{\tau }}},$
which occurs when there is a texture zero in the $(2,2)$ position, we have
the bound $\left\vert \left( U_{l}\right) _{13}\right\vert \leq \sqrt{\frac{%
m_{e}}{m_{\tau }}}.$ If, however, $\left\vert \left( U_{l}\right)
_{23}\right\vert \ll \sqrt{\frac{m_{\mu }}{m_{\tau }}}$ we have $\left(
M_{l}\right) _{22}=m_{\mu }$ and then from the determinant we have $\left(
M_{l}\right) _{13}^{2}\leq m_{e}m_{\tau }$ which again gives 
\begin{equation}
\left\vert \left( U_{l}\right) _{13}\right\vert \leq \sqrt{\frac{m_{e}}{%
m_{\tau }}}.  \label{l3}
\end{equation}%
In practice the magnitudes of $\left( U_{l}\right) _{23}$ and $\left(
U_{l}\right) _{13}$ are so small that they do not affect the mixing coming
from the neutrino sector. However $\left( U_{l}\right) _{12}$ close to the
upper bound given in eq(\ref{l1}) does give a significant contribution to
the CHOOZ angle. Its effect is considered below.

The discussion above relies on a symmetric structure relating the magnitudes
of the charged lepton mass matrix elements. If we relax this condition there
is no constraint on the magnitude of the matrix elements of $U_{l}.$ In this
case the contributions to the MNS matrix coming from the neutrino sector
should be considered as an indication of the lower bound on the $MNS$ matrix
elements, assuming there is no delicate cancellation between the
contributions of $U_{l}$ and $W.$

We turn now to a determination of the relations that follow for various form
of the Yukawa couplings.

\section{Structure of the MNS matrix\label{results}}

\subsection{ Symmetric Yukawa couplings and a single texture zero in $Y_{%
\protect\nu }^{D}$.}

\subsubsection{(1,1) texture zero}

We first consider in detail how the analysis proceeds for the case the
texture zero is in the $(1,1)$ position and both $Y_{\nu }^{D}$ and $Y_{l}^{D%
\text{ }}$ are symmetric. In the analogous case in the quark sector a $(1,1)$
texture zero leads to the remarkably successful GST relation \cite{gsto}, so
this case is particularly interesting for, if it leads to a
phenomenologically realistic prediction, it may indicate a connection
betweeen quarks and leptons.

As discussed above we are interested in the case $M_{1,2}/M_{3} 
\ll m_{2}/m_{3}%
\ $and the Majorana mass matrix, $M_{\nu }^{M}$ is diagonal and real. We
include the CP violating phases in $U_{MNS},$ i.e.we write it in the form%
\begin{equation}
U=V.diag(e^{-i\phi /2},e^{-i\phi ^{\prime }/2},1)  \label{majphase}
\end{equation}%
where $\phi $ and $\phi ^{\prime }$ are the CP violating phases and $V$ has
the form of the CKM matrix. In this case a symmetric structure in the Dirac
neutrino mass matrices and a texture zero will lead to a relation between
measurable parameters.

Following from eq(\ref{ci}) the condition $\left( Y_{\nu }^{D}\right)
_{11}=0 $ gives\footnote{%
Here and in what follows we do not include the ambiguity due to the square
roots as they can be absorbed in the unknown phases.}%
\begin{equation}
\tan z=-\sqrt{\frac{m_{2}}{m_{3}}}\frac{W_{12}^{\ast }}{W_{13}^{\ast }}
\label{tanz}
\end{equation}%
where $W$ is the matrix acting on the left-handed neutrino states needed to
diagonalise the Dirac neutrino mass matrix. To express this in terms of $%
U_{MNS}$ we use the constraints of eqs(\ref{l1},\ref{l2},\ref{l3}) to
determine $W.$ There is a residual phase ambiguity because the basis in
which the MNS matrix has the standard form can be different from the
"symmetry" basis in which the texture zero appears. This corresponds to the
simultaneous redefininition of the phase of the left- and right- handed
states such that the Dirac structure is invariant (the change in the
Majorana matrix is absorbed in a redefinition of $\phi $ and $\phi ^{\prime
} $ in eq(\ref{majphase})). With this we have $W$ $=U_{l}PU_{MNS}$ where $%
P=diag(e^{i\alpha _{1}},e^{i\alpha _{2}},e^{i\alpha _{3}}).$

From the symmetric constraint $\left( Y_{\nu }^{D}\right) _{12}=\left(
Y_{\nu }^{D}\right) _{21}$ one obtains%
\begin{equation*}
\sqrt{\frac{M_{1}}{M_{2}}}=\frac{-\tan z\sqrt{m_{2}}W_{12}^{\ast }+\sqrt{%
m_{3}}W_{13}^{\ast }}{\sqrt{m_{2}}W_{22}^{\ast }+\tan z\sqrt{m_{3}}%
W_{23}^{\ast }}.
\end{equation*}%
Substituting for $\tan z$ leads to the relation%
\begin{equation}
W_{13}^{\ast 2}+\frac{m_{2}}{m_{3}}W_{12}^{\ast 2}=-\sqrt{\frac{M_{1}}{M_{2}}}%
\sqrt{\frac{m_{2}}{m_{3}}}W_{31}\det W^{\ast }  \label{11}
\end{equation}%
where $\det W$=$e^{i\beta }$. We choose the phases of the right handed
charged leptons such that $U_{l}$ is real in the $(1,2)$ block. Then in
leading order we have $W_{ij}\simeq e^{i\alpha _{i}}U_{ij}$ except for%
\begin{equation}
W_{13}\simeq e^{i\alpha _{1}}U_{13}+e^{i\alpha _{2}}\left( U_{l}\right)
_{12}U_{23}  \label{w}
\end{equation}%
where we have written $U_{MNS}=U$. In eq(\ref{w}) we have dropped terms
involving the roots of ratios of lepton masses relative to unity. Using eq(%
\ref{w}) in eq(\ref{11}) gives 
\begin{equation}
U_{13}\equiv |U_{13}|e^{i\delta }=-e^{i(\alpha _{2}-\alpha _{1})}\left(
U_{l}\right) _{12}U_{23}\pm \sqrt{-\frac{m_{2}}{m_{3}}U_{12}^{2}-\sqrt{\frac{%
M_{1}m_{2}}{M_{2}m_{3}}}U_{31}e^{-i(\beta +2\alpha _{1})}}.  \label{1tz}
\end{equation}%
For the case of a $(1,1)$ texture zero in $\left( Y_{l}^{D}\right) _{11}$ we
have $\left( U_{l}\right) _{12}=\sqrt{\frac{m_{e}}{m_{\mu }}}$. For the case
of a texture zero in $\left( Y_{l}^{D}\right) _{12}$, $\left( U_{l}\right)
_{12}=0$. Other possibilities for a lepton texture zero or no texture zero
at all give $\left( U_{l}\right) _{12}\leq \sqrt{\frac{m_{e}}{m_{\mu }}}.$

\begin{figure}[tbp]
\centerline{
\psfig{figure=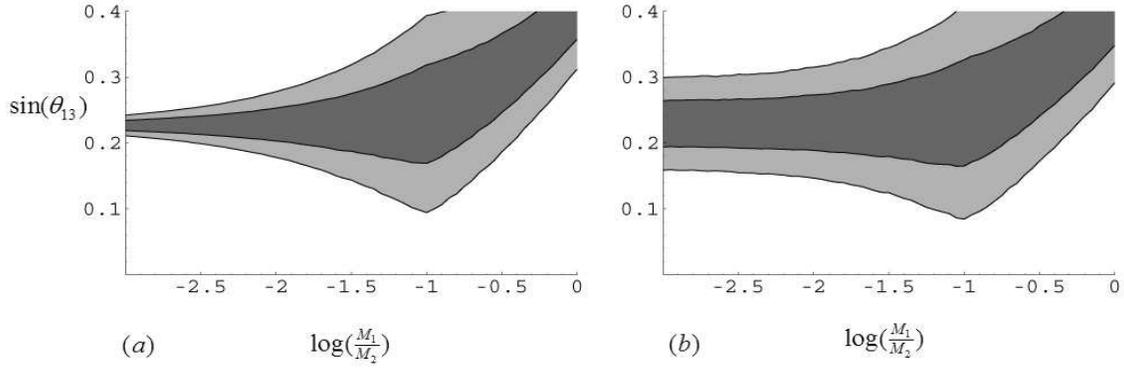,height=2.1681in,width=5.892in}}
\caption{The CHOOZ angle from a (1,1) texture zero for the limiting cases of
a simultaneous texture zero in the charged lepton mass matrix in the (a)
(1,2) and (b) (1,1) positions.}
\label{f1}
\end{figure}

The implications of eq(\ref{1tz}) for the CHOOZ angle are shown in Fig(\ref%
{f1}) for the case $\left( U_{l}\right) _{12}=0$ and $\sqrt{\frac{m_{e}}{%
m_{\mu }}}$ respectively\footnote{%
This and subsequent plots are made using the best fit points for the masses
and mixing angles of \cite{Gonzalez-Garcia:2003qf}.}. In these plots we have
chosen a random distribution of the unknown phases $\beta ,\alpha _{i}.$ One
may see there is a clustering of values within a small range with the CHOOZ
angle near the current bound, $\sin \theta _{13}<0.24$ at $3\sigma .$ This
implies that, barring an unnatural cancellation between terms, we expect a
large CHOOZ angle, in the range that would make the long baseline neutrino
factory searches for CP\ violation feasible. To quantify this we have
determined the range of the CHOOZ angle which includes 95\% of the points,
giving $\sin \theta _{13}>0.1$ over the whole range of $M_{1}/M_{2}.$

\begin{figure}[tbp]
\centerline{
\psfig{figure=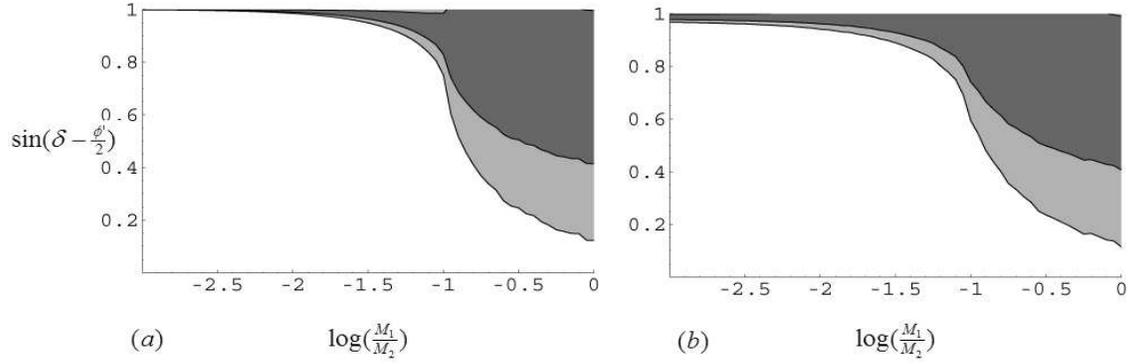,height=2.1681in,width=5.892in}}
\caption{The $\protect\nu 0 \protect\beta \protect\beta $ CP violating phase
from a (1,1) texture zero for the limiting cases of a simultaneous texture
zero in the charged lepton mass matrix in the (a) (1,2) and (b) (1,1)
positions.}
\label{f2}
\end{figure}

In Fig(\ref{f2}) we plot the distribution for the CP violating phase
combination $\sin (\delta -\phi ^{\prime }/2)$. This is the CP\ violating
phase relevant to neutrinoless double beta decay. We see that $\sin (\delta
-\phi ^{\prime }/2)$ clusters near its maximal value. In this case the 95\%
cutoff implies $\sin (\delta -\phi ^{\prime }/2)>0.4.$

Finally we determine the implications of our results for thermal
leptogenesis, assuming that the lightest Majorana state dominates\cite%
{Buchmuller:2003gz}. In this case the asymmetry is given by 
\begin{eqnarray*}
\epsilon &\simeq&-\frac{3}{8\pi }\frac{M_{1}}{v^{2}}\frac{{\rm Im}(\cos ^{2}z%
\text{ }m_{2}^{2}+\sin ^{2}z\text{ }m_{3}^{2})}{m_{2}\left\vert \cos
^{2}z\right\vert +m_{3}\left\vert \sin ^{2}z\right\vert } \\
&=&-\frac{3}{8\pi }\frac{M_{1}}{v^{2}}\frac{(m_{3}^{2}-m_{2}^{2}){\rm Im}%
(\sin ^{2}z)}{m_{2}\left\vert \cos ^{2}z\right\vert +m_{3}\left\vert \sin
^{2}z\right\vert }
\end{eqnarray*}%
Since $|\epsilon _{\max }|=\frac{3}{8\pi} \frac{M_{1}m_{3}}
{\langle H^{0}\rangle^2}$
$\cite{Davidson:2002qv},$ we
have%
\begin{equation*}
\frac{\epsilon }{|\epsilon _{\max }|}\simeq -\frac{{\rm Im}(\sin ^{2}z)}{%
\left\vert \sin ^{2}z\right\vert +\frac{m_{2}}{m_{3}}\left\vert \cos
^{2}z\right\vert }
\end{equation*}

\begin{figure}[t]
\centerline{
\psfig{figure=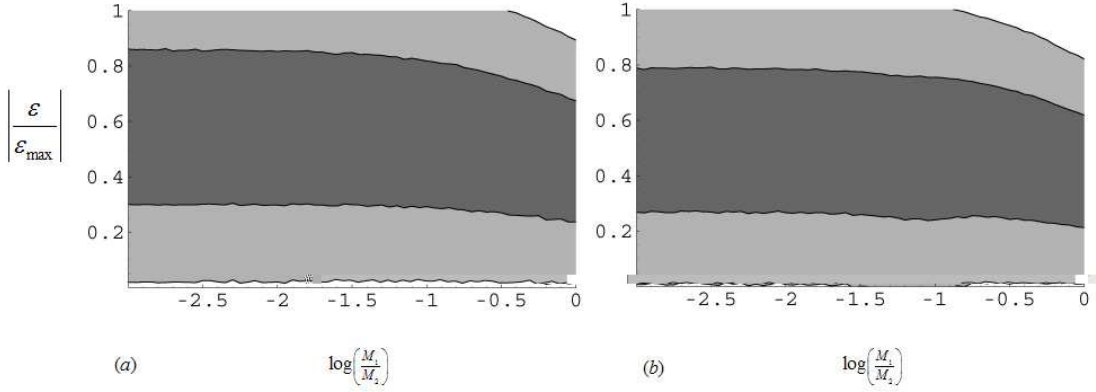,width=5.892in}}
\caption{The CP\ asymmetry compared to the maximal value in thermal
leptogenesis from a (1,1) texture zero for the limiting cases of a
simultaneous texture zero in the charged lepton mass matrix in the (a) (1,2)
and (b) (1,1) positions.}
\label{f3}
\end{figure}

Note that $\epsilon $ depends only on $\tan z$. The dependence of $\tan z$
on low energy phases may be read from eq(\ref{tanz}) showing which
combination is relevant for leptogenesis. The magnitude of $\epsilon
/\epsilon _{\max }$ is plotted in Fig(\ref{f3}). Note that, if we ignore the charged
lepton contribution coming from a nontrivial $U_l$, a $(1,1)$ texture zero with an hierarchical Majorana
mass spectrum gives the same value for the CP violating phase in double beta
decay as the CP\ violating phase determining the lepton asymmetry in
leptogenesis \cite{King:2002qh}. This explains the correlation seen between
the plots of Figs(\ref{f3}), although note that in Figs(\ref{f3}b) a
significant charged lepton contribution has been added.

\begin{figure}[tbp]
\centerline{
\psfig{figure=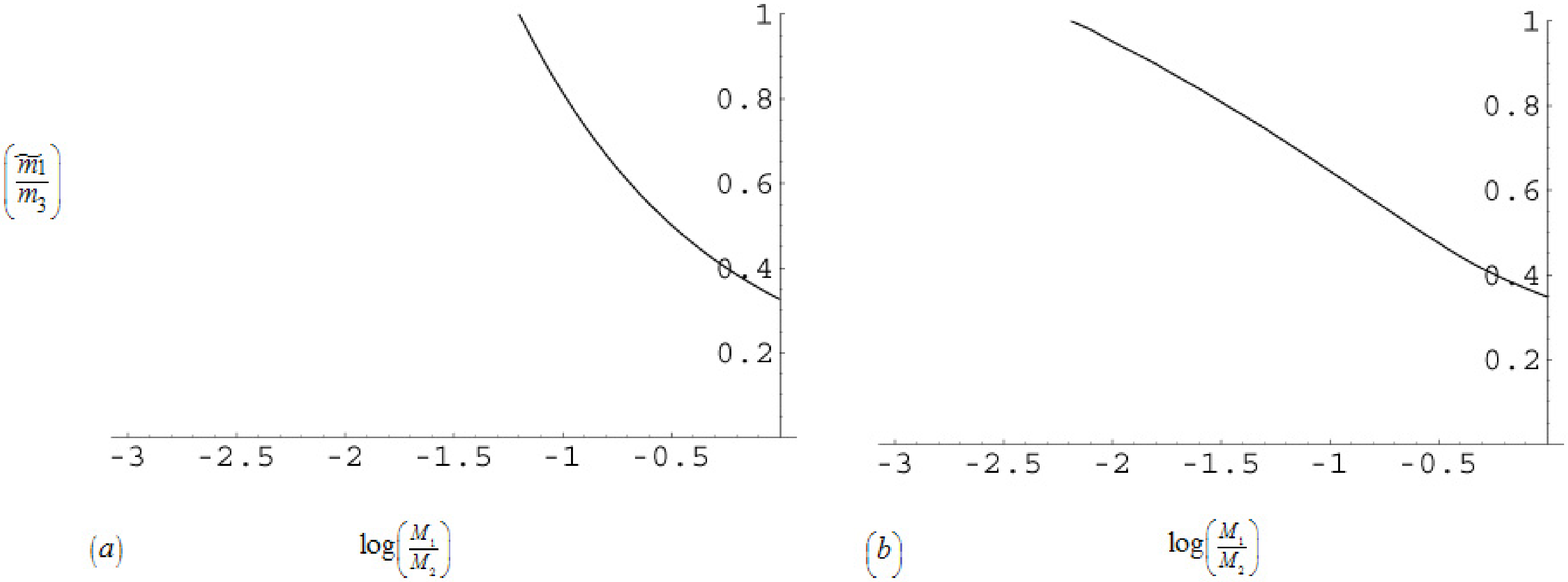,height=2.1681in,width=5.892in}}
\caption{A plot of the lower bound of $\widetilde{m}_{1}/m_{3}$ versus $\log
M_{1}/M_{2\text{ }}$ for the case of a (1,1) texture zero.}
\label{f4}
\end{figure}

Whether this asymmetry can lead to the observed baryon asymmetry depends on
the subsequent washout. This is characterised by the parameter $\widetilde{m}%
_{1}$ \cite{Plumacher:1996kc}$.$ It is given by 
\begin{equation*}
\widetilde{m}_{1}=m_{2}\left\vert \cos ^{2}z\right\vert +m_{3}\left\vert
\sin ^{2}z\right\vert
\end{equation*}%
For the case of a (1,1) texture zero the value of $\widetilde{m}_{1}$ is
given in Fig(\ref{f4}). In the whole region of parameter space 
$\widetilde{m}_{1}\gg m_{2}$ and so the washout will reduce the baryon asymmetry
below the observed value unless $M_{1}$ is very large \cite%
{Chankowski:2003rr}. In the case of SUGRA this implies a reheat temperature
above the gravitino abundance bound implying that in this case thermal
leptogenesis cannot work. However in other supersymmetry breaking mediation
scenarios, such as gauge mediation, the gravitino is much lighter and a
heavier $M_{1}$ is consistent with the gravitino bound.

\subsubsection{A single texture zero in the (1,2), (1,3), (2,2) or (2,3)
positions}

It is straightforward to apply the analysis just discussed to the other
possible positions for a single texture zero in the Dirac neutrino matrix.
The results are presented in Table \ref{table1}. Note that, unlike the case
for a $(1,1)$ texture zero, the prediction for $\tan z$ is in terms of the
measured large $MNS$ matrix elements. As a result one obtains a definite
prediction for leptogenesis which is also given in the Table. For the case
of $(1,2)$ and $(1,3)$ texture zeros we see that $\tan z$ is suppressed by $%
\sqrt{\frac{m_{2}}{m_{3}}}$ which leads to a near maximal form for $%
\frac{\epsilon }{\epsilon _{\max }}.$ The bound on $
\widetilde{m}_{1}$ is only mildly stronger than the absolute bound
$\widetilde{m}_{1}\ge m_{2}$, so the washout effects are expected
to be less efficient than in the $(1,1)$ texture zero case.
For the case of the $(2,2)$ and $(2,3)$ texture zeros $\tan z$ is enhanced
by $\sqrt{\frac{m_{3}}{m_{2}}}$ which leads to a 
$\frac{m_{2}}{m_{3}}$
suppression in $\frac{\epsilon }{\epsilon _{\max }}.$ The bound on $%
\widetilde{m}_{1}$ in this case is comparable to the one for 
a $(1,1)$ texture zero but is independent of $M_{1}/M_{2}.$
As a result baryogenesis through thermal leptogenesis 
will not proceed in these cases either.

\begin{figure}[tbp]
\centerline{
\psfig{figure=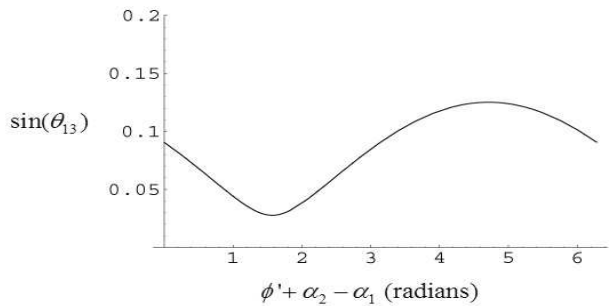,height=1.7634in,width=3.2863in}}
\caption{The CHOOZ angle for the (1,2) texture zero plotted against the
unknown phase. }
\label{f5}
\end{figure}

\begin{figure}[tbp]
\centerline{
\psfig{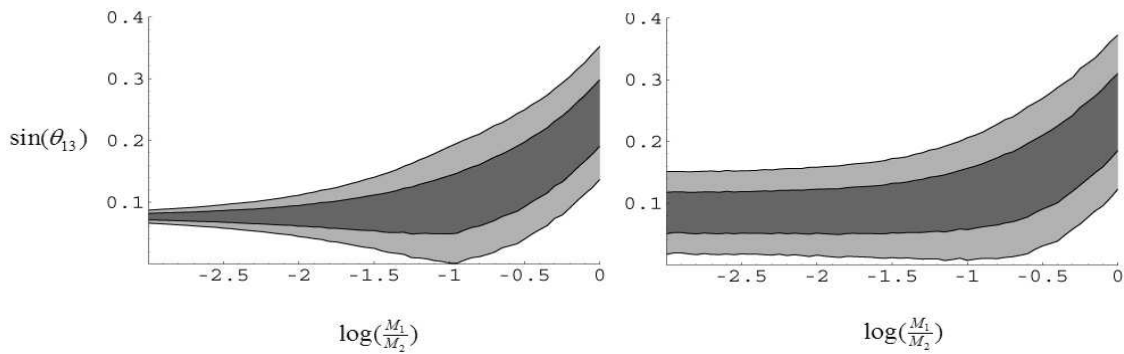}}
\caption{The CHOOZ angle from a (1,3) texture zero for the limiting cases of
a simultaneous texture zero in the charged lepton mass matrix in the (a)
(1,2) and (b) (1,1) positions. }
\label{f6}
\end{figure}

For the case of the $(1,2)$ texture zero the prediction for the CHOOZ angle
depends only on unknown phases with the distribution is shown in Fig(\ref{f5}%
). For a (1,3) texture zero the CHOOZ angle also depends on the ratio $%
M_{1}/M_{2}$ as in the previous cases. This is plotted in Fig(\ref{f6}). In
both cases $\theta _{13}$ is predicted to be large, although the $95\%$
lower range is smaller than that found for the $(1,1)$ texture zero case.

For the case of the $(2,2)$ and $(2,3)$ texture zeros one obtains a relation
between the large elements of the MNS matrix. From this one may extract a
relation between the phases and a prediction for $M_{1}/M_{2}.$
Unfortunately these do not lead to a relation between measureable
parameters, although the constraint that $M_{1}/M_{2}\simeq m_{2}/m_{3}$ may
be of interest in model building.

\begin{table}[tbp] \centering%
\begin{tabular}{|c|c|c|c|c|}
\hline
\textbf{Texture } & $tan$ $z$ & $\frac{\epsilon }{\epsilon _{\max }}$ & $%
\widetilde{m}_{1}$ & MNS relation \\ 
\textbf{zero} &  &  &  &  \\ \hline
$\mathbf{(1,1)}$ & $\sqrt{\frac{m_{2}}{m_{3}}}\frac{W_{12}^{\ast }}{%
W_{13}^{\ast }}$ & $see$ $text$ & $\gg m_{2}$ & $\mathbf{U}_{13}=-\chi
e^{i(\alpha _{2}-\alpha _{1})}\sqrt{\frac{m_{e}}{m_{\mu }}}U_{23}$ \\ 
&  &  &  & $\pm \sqrt{-\frac{m_{2}}{m_{3}}U_{12}^{2}+\sqrt{\frac{M_{1}m_{2}}{%
M_{2}m_{3}}}U_{31}e^{-i(\beta +2\alpha _{1})}}$ \\ \hline
$\mathbf{(1,2)}$ & $\sqrt{\frac{m_{2}}{m_{3}}}\frac{U_{22}^{\ast }}{%
U_{23}^{\ast }}$ & -$\frac{\sin \phi ^{\prime }c_{12}^{2}}{1+c_{12}^{2}}$ & 
$\simeq \frac{m_{2}(1+c_{_{12}}^{2})}{1+\frac{m_{2}}{m_{3}}%
c_{_{12}}^{2}}$ & $\mathbf{U%
}_{13}=-\chi e^{i(\alpha _{2}-\alpha _{1})}\sqrt{\frac{m_{e}}{m_{\mu }}}%
U_{23}$ \\ 
&  &  &  & $-\frac{m_{2}}{m_{3}}\frac{U_{12}U_{22}}{U_{23}}$ \\ \hline
$\mathbf{(1,3)}$ & $\sqrt{\frac{m_{2}}{m_{3}}}\frac{U_{32}^{\ast }}{%
U_{33}^{\ast }}$ & -$\frac{\sin \phi ^{\prime }c_{12}^{2}}{1+c_{12}^{2}}$ & 
$\simeq \frac{m_{2}(1+c_{_{12}}^{2})}{1+\frac{m_{2}}{m_{3}}%
c_{_{12}}^{2}}$ & $\mathbf{U%
}_{13}=-\chi e^{i(\alpha _{2}-\alpha _{1})}\sqrt{\frac{m_{e}}{m_{\mu }}}%
U_{23}-\frac{m_{2}}{m_{3}}\frac{U_{12}U_{32}}{U_{33}}$ \\ 
&  &  &  & $+e^{i(\beta -2\alpha _{1}-\alpha _{3})}\sqrt{\frac{M_{1}}{M_{2}}}%
\sqrt{\frac{m_{2}}{m_{3}}}\frac{U_{11}^{\ast }}{U_{33}}$ \\ \hline
$\mathbf{(2,2)}$ & $\sqrt{\frac{m_{3}}{m_{2}}}\frac{U_{23}^{\ast }}{%
U_{22}^{\ast }}$ & $\frac{\sin \phi ^{\prime }m_{2}}{m_{3}}c_{12}^{2}$ & 
$\simeq \frac{m_{3}}{1+\frac{m_{2}}{m_{3}}c_{_{12}}^{2}}$ & 
$U_{31}=e^{i(\beta -2\alpha _{2}-\alpha _{3})}\sqrt{\frac{%
M_{1}}{M_{2}}}\sqrt{\frac{m_{3}}{m_{2}}}(U_{23}^{\ast 2}+\frac{m_{2}}{m_{3}}%
U_{22}^{\ast 2})$ \\ \hline
$\mathbf{(2,3)}$ & $\sqrt{\frac{m_{3}}{m_{2}}}\frac{U_{33}^{\ast }}{%
U_{32}^{\ast }}$ & $\frac{\sin \phi ^{\prime }m_{2}}{m_{3}}c_{12}^{2}$ & 
$\simeq \frac{m_{3}}{1+\frac{m_{2}}{m_{3}}c_{_{12}}^{2}}$ & 
$U_{21}=e^{i(\beta -2\alpha _{2}-\alpha _{3})}\sqrt{\frac{%
M_{1}}{M_{2}}}\sqrt{\frac{m_{3}}{m_{2}}}(U_{23}^{\ast }U_{33}^{\ast }+\frac{%
m_{2}}{m_{3}}U_{22}^{\ast }U_{32}^{\ast })$ \\ \hline
\end{tabular}%
\caption{The constraints following from a symmetric mass matrix and a single texture zero . 
$\chi$ is 1 for a (1,1) texture zero in the charged lepton sector and 0 for a (1,2) texture zero.
If there is no lepton texture zero $\chi$ lies between these limiting cases. $c_{12}$ is 
$cos(\theta_{12})$.\label{table1}}%
\end{table}%

\subsection{The case of two texture zeros}

For two texture zeros one obtains a prediction even without imposing the
symmetric constraint. There are fifteen ways of assigning two texture zeros
to the first two rows of the Dirac neutrino mass matrix (the third row plays
no role in the case the third Majorana neutrino is anomalously heavy). All
but five lead to inconsistent results; below we discuss only the viable
choices.

From Table \ref{table1} we may readily solve the constraint following from
equating the two forms for $\tan z$ that follow from $(1,1)$ and $(2,2)$
texture zeros. This gives the prediction for $U_{13}$ given in Table \ref%
{table2}. One may see it is identical to the prediction (\textit{c.f.}
Figure \ref{f5}) obtained for a single texture zero in the $(1,2)$ position
with the symmetric condition imposed although in this case we have not
imposed this condition. If one further imposes the condition that the matrix
is symmetrical one also obtains the prediction for $U_{13}$ given in eq(\ref%
{1tz}). Equating these results fixes one combination of the phases (which
does not lead to new relations between measurable phases) and fixes the
ratio $M_{1}/M_{2}\simeq m_{2}/m_{3}.$ The prediction for $\frac{\epsilon }{%
\epsilon _{\max }}$ is as given in Table \ref{table1} for the $(2,2)$
texture zero case.

\begin{figure}[tbp]
\centerline{
\psfig{figure=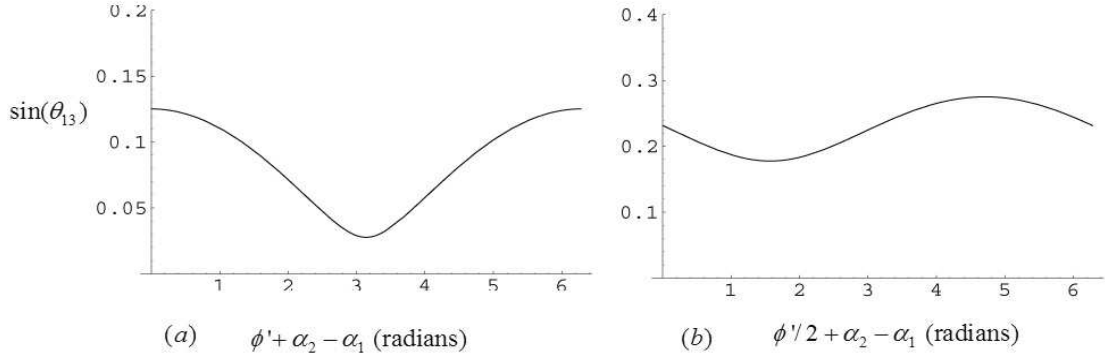,height=2.1681in,width=5.892in}}
\caption{The prediction for the CHOOZ\ \ angle for the two texture zero
cases\ : (a) $(1,1)$ and $(2,3)$ or $(1,3)$ and $(2,1)$ (b) $(1,1)$ and $%
(2,1).$ The plot is for the $\protect\chi =1$ case and is plotted againt the
relative phase between the two terms appearing in Table \protect\ref{table2}%
. }
\label{f7}
\end{figure}

The remaining possibilities are given in Table \ref{table2}. The prediction
for the CHOOZ angle is approximately the same for the $(1,1)$ and $(2,3)$ or
the $(1,3)$ and $(2,1)$ cases and is shown in Figure \ref{f7}(a). The
remaining case with a $(1,1)$ and a $(2,1)$ texture zero is shown in Figure %
\ref{f7}(b).

For the case of $(1,1)$ and $(2,3)$ texture zeros one again needs $%
M_{1}/M_{2}\simeq m_{2}/m_{3}$ if one requires the Dirac mass matrix be
symmetric$.$ For the last two cases there is no solution if one additionally
imposes the condition the Dirac mass matrix be symmetric. In all cases the
prediction for $\frac{\epsilon }{\epsilon _{\max }}$ is as given in Table %
\ref{table1} for the appropriate texture zero. This follows because the
prediction comes from the constraint on $\tan z$ only and does not require
the symmetric condition.

\begin{table}[tbp] \centering%
\begin{tabular}{|c|c|}
\hline
\textbf{Texture zero} & $\mathbf{U}_{13}$ \\ \hline
$(1,1)$ and $(2,2)$ & $\pm \frac{m_{2}}{m_{3}}\frac{U_{12}U_{22}}{U_{23}}%
-\chi e^{i(\alpha _{2}-\alpha _{1})}\sqrt{\frac{m_{e}}{m_{\mu }}}U_{23}$ \\ 
\hline
$(1,1)$ and $(2,3)$ & $\pm \frac{m_{2}}{m_{3}}\frac{U_{12}U_{32}}{U_{33}}%
-\chi e^{i(\alpha _{2}-\alpha _{1})}\sqrt{\frac{m_{e}}{m_{\mu }}}U_{23}$ \\ 
\hline
$(1,1)$ and $(2,1)$ & $\pm \sqrt{\frac{m_{2}}{m_{3}}}U_{12}-\chi e^{i(\alpha
_{2}-\alpha _{1})}\sqrt{\frac{m_{e}}{m_{\mu }}}U_{23}$ \\ \hline
$(1,3)$ and $(2,1)$ & $\pm \frac{m_{2}}{m_{3}}\frac{U_{12}U_{32}}{U_{33}}%
-\chi e^{i(\alpha _{2}-\alpha _{1})}\sqrt{\frac{m_{e}}{m_{\mu }}}U_{23}$ \\ 
\hline
\end{tabular}%
\caption{The constraints following from two texture zeros. Only those cases shown are consistent apart from the (1,2), (2,1) case which has already been discussed when considering symmetric textures. Also shown are the additional 
constraints following from imposing a symmetric structure for the two cases this is consistent. $\chi$ is 0 for a (1,1) texture zero in the charged lepton sector and 0 for a (1,2) texture zero.
 For no lepton texture zero $\chi$ is between these limiting cases. \label{table2}}%
\end{table}%

\section{Summary and Conclusions\label{summary}}

The combination of the see-saw mechanism, an hierarchical structure for the
Majorana mass matrix and a combination of texture zeros and/or a symmetrical
form for the moduli of the mass matrix elements leads to relations amongst
observable properties of neutrinos. In this paper we have determined these
predictions in a model independent way.

The case of a $(1,1)$ texture zero is of particular interest because, in the
quark sector, it leads to a relation in excellent agreement with experiment.
In the neutrino case the equivalent $(1,1)$ texture zero leads to a
prediction for the CHOOZ angle that is close to the present limit and a near
maximal CP\ violating phase relevant to thermal leptogenesis and to $\nu
0\beta \beta .$ For the $(1,1)$ texture zero, thermal 
leptogenesis cannot give rise to acceptable
baryogenesis while satisfying the gravitino bounds on the reheat
temperature. Therefore, an acceptable range of baryogenesis is
only possible if the gravitino constraints are relaxed, for example
in theories in which the supersymmetry breaking occurs at a lower scale.

In the case that the texture zero appears in the $(1,2)$ or $(1,3)$
positions the CHOOZ angle is still predicted to be large, encouraging for
long baseline CP\ violation studies. Furthermore, in these cases
washout effects after thermal leptogenesis are not too efficient
and could allow for adequate baryogenesis.
The case of $(2,2)$ and $(2,3)$ texture zeros does not lead to
phenomenologically interesting relations. However there are five
viable cases in which two texture zeros can be present. In these cases a
large CHOOZ angle is again predicted.

The determination of the parameters involved in the see-saw \ mechanism is
an illdefined problem due to the large number of parameters relative to
measureables. The best hope is that the system has a high degree of
symmetry, reducing the number of parameters. Our analysis has explored a
particularly promising possibility suggested by the structure observed in
the quark sector in which the Dirac masses have one (or more) texture
zero(s) and the magnitude of the mass matrix elements may be symmetric. In
addition we have assumed an hierarchical structure for the Majorana matrix,
motivated by the fact this can readily explain the large neutrino mixing
angles while having a relation between quark and lepton Dirac masses. Such a
structure for the Dirac and Majorana masses can be derived from an
underlying family symmetry\cite{skggr} and, if the resultant predictions for
neutrino properties should be confirmed, it would provide strong support for
such an underlying symmetry organising the fermion mass matrices.

\begin{center}
\textbf{Acknowledgement}
\end{center}

This work was partly supported by the EU network, \textquotedblleft
Physics Across the
Present Energy Frontier\textquotedblright\ HPRV-CT-2000-00148 and the PPARC rolling grant PPA/G/O/2002/00479.

\end{document}